# THE KRAMERS-KRONIG RELATIONS
# FOR LIGHT MOVING BACKWARDS IN TIME

**Evangelos Chaliasos**

365 Thebes Street

GR-12241 Aegaleo

Athens, Greece

*Abstract*

The recent discovery of light moving *backwards* in time, when it propagates in a suitable dispersive medium, obliges us to reexamine the Kramers-Kronig relations. In their usual form, they are dealing with usual light (moving forward in time), and they are a consequence of causality. In a similar fashion, we derive the appropriate form of these relations for light moving *backwards* in time, starting, instead of causality, from a principle of *anti-causality,* appropriate to such a light.

**1. Introduction**

In a former paper [1] the author explained how we can receive light before its emission [2, 3], if we accept that this light moves backwards in time following the so-named *principle of anticausality,* that is that its radius vector is *time-like,* but obeying the condition (in one dimension) $ct - x < 0$ & $ct + x < 0$ (anti-causality), instead of $ct - x > 0$ & $ct + x > 0$ (causality), obeyed by usual light. Wang, Kuzmich & Dogariu [2] found a negative group-velocity index of refraction starting from a formula (eq. (1) of [2]) resulting from the Kramers-Kronig relations (for a derivation of them see [4, 5])

(and using their experimental values of the parameters entered), which are based on the assumption c t - x > 0 ( x > 0 ) (causality). And they concluded thus that causality is not violated, contrary to the fact that their pulse obviously moved backwards in time. But in order for the Kramers-Kronig relations to hold (in a modified of course form) it is not necessary to start from the relation c t - x > 0 ( x > 0 ) (causality). Thus, it will be found that it is sufficient to start from the more general relation

$c^2 t^2 - x^2 > 0,$

which may be called the *generalized causality principle.* In fact, as we will show in what follows, the appropriate starting point for light moving backwards in time is the relations c t - x < 0 & c t + x < 0 (anti-causality), from which we can again take a modified form of the Kramers-Kronig relations, which remain essentially the same, and thus explaining the apparently strange effect discovered by the above mentioned authors last year.

## 2. Establishment of analyticity of n(ω)

The refractive index n of an optical medium is defined as the fraction of the velocity of propagation of light in vacuum over the phase velocity inside the medium,
$n = c/v_p = ck/\omega$
(where k is the wave number and ω is the cyclic frequency), and it is a physical quantity of basic significance. From the relation
$k = n\omega/c,$
it results for the group velocity by derivation that
$v_g = d\omega/dk = c/(n + \omega \, dn/d\omega).$ (1)
The refractive index is considered as a function of the (cyclic) frequency ω, and it will be proven, in what follows, that, as a consequence of anti-causality, n(ω) is an analytic function of ω in the *lower* complex half-plane.

In order to see this, let us consider for simplicity an optically homogeneous medium, with refractive index n(ω), in one dimension, inside of which an "electromagnetic disturbance" of the form E ( x , 0 ) = δ( x ) ( x > 0 ) takes place, in time t = 0 . If we resolve the disturbance according to Fourier,

$$\delta(x) = \frac{1}{2\pi} \int_{-\infty}^{+\infty} e^{ikx} dk, \qquad (2)$$

and if we take in mind the wave propagation, the disturbance above, in time t, takes the form

$$E(x,t) = \frac{1}{2\pi} \int_{-\infty}^{+\infty} e^{-i(\omega t - kx)} \frac{dk}{d\omega} d\omega, \qquad (3)$$

where $\omega t - kx > 0$, if we accept that the wave moves either forward ( $t > 0$ ) or *backwards* ( $t < 0$ ) in time. So, we find

$$E(x,t) = \frac{1}{2\pi} \int_{-\infty}^{+\infty} e^{-i\omega\left[t - \frac{n(\omega)}{c}x\right]} \frac{dk}{d\omega} d\omega. \qquad (4)$$

In the complex plane then (see fig. 1) we find[*)]

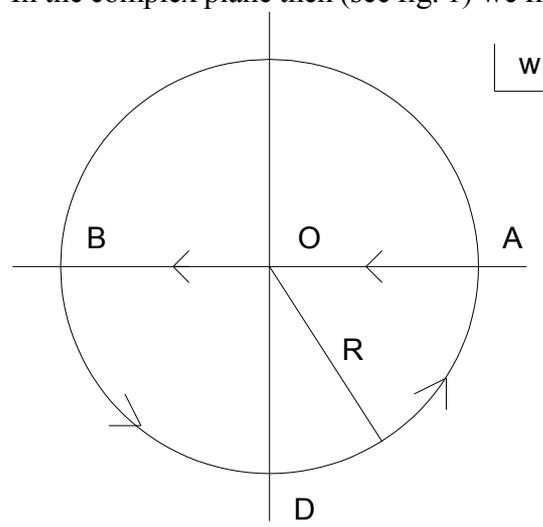

ABDA is C

BDA is $C_R$

Fig. 1

$$-E(x,t) = \lim_{R \to \infty} \left\{ \frac{1}{2\pi} \int_C e^{-i\omega\left[t - \frac{n(\omega)}{c}x\right]} \frac{dk}{d\omega} d\omega - \frac{1}{2\pi} \int_{C_R} e^{-i\omega\left[t - \frac{n(\omega)}{c}x\right]} \frac{dk}{d\omega} d\omega \right\}. \qquad (5)$$

Now, if n(ω) is analytic in the *lower* half-plane and
$n(\omega) \to -1$ for $|\omega| \to \infty$,

---

[*)] The reason for which we are considering ω with negative imaginary part is for the medium to be absorbing, as it will be clear from what follows. Even the real part of ω must here be understood as opposite in sign as compared to the corresponding ω for motion forward in time. This is the case because for motion backwards in time the period of the wave-generating vibration must be taken negative.

(**) the closed integral vanishes, so that we are left with

$$E(x,t) = \lim_{R \to \infty} \frac{1}{2\pi} \int_{C_R} e^{-i\omega'\left(t+\frac{x}{c}\right)+\omega''\left(t+\frac{x}{c}\right)} \frac{dk}{d\omega} d\omega, \tag{6}$$

where
$\omega' = \text{Re}\,\omega$ & $\omega'' = \text{Im}\,\omega$.

When $t + x/c > 0$, the limit above equals zero, and therefore $E(x,t) = 0$. In the case that n(ω) was not analytic in the *lower* half-plane, we would have $E(x,t) \neq 0$ for $t + x/c > 0$ (since then the closed integral in (5) would not vanish), a fact which would violate the principle of anti-causality, which we have assumed to hold since we are dealing with light moving *backwards* in time. Thus n(ω) is an analytic function of ω in the *lower* half-plane.

### 3. Derivation of the relations

Together with n(ω), the expression n(ω) + 1 is also analytic in the lower half-plane. Therefore the following Cauchy relation will be valid

$$n(\omega) + 1 = \frac{1}{2\pi} \int \frac{n(\omega')+1}{\omega'-\omega} d\omega'. \tag{7}$$

We consider that, the integral is taken along an infinite semi-circle with base the real axis, *except* from a small semi-circle, which excludes the singularity at $\omega' = \omega$ from the perimeter (better see fig. 2).

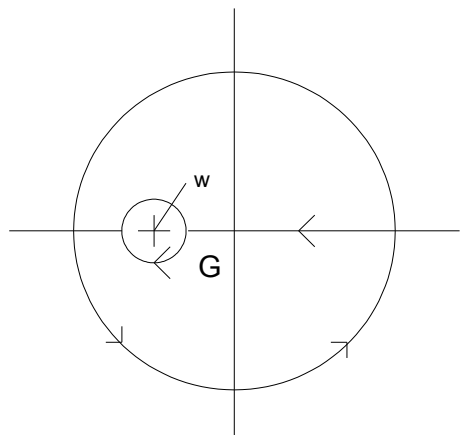

Fig. 2. The new closed curve is now named C.

With the assumption that
$\lim_{\omega \to \infty}[n(\omega)+1] = 0,$
and the property that

---
(**) In order for the phase velocity to become v → c at infinity (we remind that x > 0), since t < 0 in our case.

$$n(-\omega) = n^*(\omega),$$

(***) we have

$$\int_C \frac{n(\omega')+1}{\omega'-\omega}d\omega' = 0, \qquad (8)$$

or

$$\int_{+\infty}^{\omega} + \int_{\omega}^{-\infty} + \int_G + \int_{C_R} = 0 \qquad (9)$$

But because R $\to\infty$, we have

$$\int_{C_R} = 0, \qquad (10)$$

so that we obtain

$$\int_{+\infty}^{0}\frac{n(\omega')+1}{\omega'-\omega}d\omega' + \int_{0}^{-\infty}\frac{n(\omega')+1}{\omega'-\omega}d\omega' + \int_G \frac{n(\omega')+1}{\omega'-\omega}d\omega' = 0. \qquad (11)$$

If we introduce (because we essentially take limits) the Cauchy principal values for the first two integrals, we find

$$P\int_0^{-\infty}\frac{\operatorname{Re}n(\omega') - i\operatorname{Im}n(\omega')+1}{-\omega'-\omega}d\omega' + P\int_0^{-\infty}\frac{\operatorname{Re}n(\omega') + i\operatorname{Im}n(\omega')+1}{\omega'-\omega}d\omega' +$$

$$+[\operatorname{Re}n(\omega) + i\operatorname{Im}n(\omega)+1]\int_G \frac{d\omega'}{\omega'-\omega} = 0. \qquad (12)$$

But

$$\int_G \frac{d\omega'}{\omega'-\omega} = -\pi i. \qquad (13)$$

Thus, we find that

$$2P\int_0^{-\infty}\frac{\omega[\operatorname{Re}n(\omega')+1] + i\omega'\operatorname{Im}n(\omega')}{\omega'^2-\omega^2}d\omega' - \pi i[\operatorname{Re}n(\omega) + i\operatorname{Im}n(\omega)+1] = 0 \qquad (14)$$

Equating both imaginary and real part of the left hand side of (14) to zero, we finally find respectively

$$\operatorname{Re}n(\omega)+1 = \frac{2}{\pi}P\int_0^{-\infty}\frac{\omega'\operatorname{Im}n(\omega')}{\omega'^2-\omega^2}d\omega' \qquad (15)$$

and

$$\operatorname{Im}n(\omega) = -\frac{2\omega}{\pi}P\int_0^{-\infty}\frac{\operatorname{Re}n(\omega')+1}{\omega'^2-\omega^2}d\omega' \qquad (16)$$

These relations (15) & (16) constitute the dispersion relations of Kramers-Kronig in the case of light moving *backwards* in time.

### 4. Comments

---

(***) In order for E ( x , t ) to take real values.

Like the normal Kramers-Kronig relations, equations (15) and (16) are of fundamental importance because, in order to prove them, special dynamical assumptions are not required. Only locality and *anti-causality* are needed. If we take in mind that now ω' and ω are of the opposite sign compared to ω' and ω (respectively) of the normal case, and also that the same happens concerning n, while
$$-1 = \lim_{\omega \to \infty} n(\omega)$$
(versus
$$+1 = \lim_{\omega \to \infty} n(\omega)$$
of the normal case), we see that the Kramers-Kronig relations essentially do not change. This is why Wang, Kuzmich & Dogariu [2] found the correct value for n, that is n < - 1 (< 0) , eventhough the anti-causality principle rather than the causality principle obviously holds (since undoubtedly the light pulse's motion is backwards in time).

## 5. Application

Ending the paper, we mention by the way that, as we know, Imn(ω') represents the absorption coefficient of the propagation medium. In the case that the absorption is due to an ideal resonance of the form
$$\mathrm{Im}\, n(\omega') = g\pi\delta(\omega' - \omega_0), \tag{17}$$
with $\omega_0$ the resonance frequency and δ the delta function, the relation (15) gives

$$\mathrm{Re}\, n(\omega) = -1 + 2g \int_0^{-\infty} \frac{\omega'\delta(\omega' - \omega_0)}{\omega'^2 - \omega^2} d\omega', \tag{18}$$
or
$$\mathrm{Re}\, n(\omega) = -1 - 2g \frac{\omega_0}{\omega_0^2 - \omega^2}, \tag{19}$$

that is it is now negative, for | g| not too large. But of course now g < 0 for extinction and g > 0 for emission.